\documentclass[aps,pre,twocolumn]{revtex4-1}
\pdfoutput=1

\usepackage{graphicx}
\graphicspath{{images/}}
\usepackage{color}

\begin{document}

  \title{Community detection in networks using self-avoiding random walks}

  \author{Guilherme de Guzzi Bagnato}\email{guilherme.bagnato@usp.br}
  \author{Jos\'{e} Ricardo Furlan Ronqui}
  \author{Gonzalo Travieso}\email{gonzalo@ifsc.usp.br}
  \affiliation{Instituto de F\'{\i}sica de S\~ao Carlos, Universidade
    de S\~ao Paulo, CP 369, 13560-970, S\~ao Carlos, SP, Brazil}

  \date{\today}

  \begin{abstract}

    Different kinds of random walks have proven to be useful in the study of  structural properties of complex networks. Among them, the restricted dynamics of self-avoiding random walks (SAW), which visit only at most once each vertex in the same walk, has been successfully used in network exploration. The detection of communities of strongly connected vertices in networks remains an open problem, despite its importance, due to the high computational complexity of the associated optimization problem and the lack of a unique formal definition of communities. In this work, we propose a SAW-based method to extract the community distribution of a network and show that it achieves high modularity scores, specially for real-world networks. We combine SAW with principal component analysis to define the dissimilarity measure to be used for agglomerative hierarchical clustering. To evaluate the performance of this method we compare it with four popular methods for community detection: Girvan-Newman, Fastgreedy, Walktrap and Infomap using two types of synthetic networks and six well-known real-world cases.

  \end{abstract}

  \maketitle

  \section{Introduction}

   In the past few years, many dynamical processes such as percolation \cite{Stauffer1971}, synchronization \cite{Arenas2006} and epidemic spreading \cite{Pastor-Satorras2001} have been studied in a wide variety of complex networks. Among these processes, random walks have proven to be a flexible tool to characterize and explore networks. In a random walk, a walker visits one vertex per step chosen randomly among all neighbors of the current vertex. Although simple, this process has the advantage of using only local information of the network, making it convenient when few properties of the system as a whole are known. Several properties of random walks on complex networks have been studied as, for example, scaling behavior in small world networks \cite{Lahtinen2001,Almaas2003,Lahtinen2001}, first passage time \cite{Noh2004,Masuda2004,Tejedor2009}, characteristics of this dynamic in directed networks \cite{Tadi2001,Comin2014,Aksoy2016} as well as the effect of finite memory \cite{Lambiotte2014,Weng2014}.

   Many alternative types of random walks have been proposed to optimize topological analyses \cite{Yang2005,Costa2007,LopezMillan2011,Marshak2016,Guo2016,Kim2016}.  Among these approaches, the self-avoiding walk (SAW) was shown to be more efficient in the exploration and navigation of different network structures than the traditional walker. In the SAW, the walkers cannot return to an already visited vertex, forcing them to find a new viable path through unvisited vertices. If no new vertices are available, the walk ends. Because the walker retains a memory of the path traveled, a general analytic solution is not trivial. However, some theoretical efforts yielded interesting results for small-world networks \cite{Herrero2003}, Erd\H{o}s-R\'{e}nyi \cite{Tishby2016} and scale-free \cite{Herrero2005-0,Herrero2005-1} topologies.

   One feature of complex networks that has been subject of research in several fields such as physics, biology and economy is the presence of a community structure: groups of vertices densely connected to each other and (comparatively) sparsely connected with the rest of the network~\cite{Fortunato2010}.  In recent years, a variety of methods for identifying those groups were developed using dynamical process based on structural properties, including random walks \cite{Zhou2003,Rosvall29012008,Steinhaeuser2010,Xin2016,Zhu2016}. In order to find the best community distribution, such algorithms as spectral methods \cite{Newman06062006,Danila2015} and extremal optimization~\cite{Arenas2005} frequently work by optimizing the quality function known as modularity \cite{Newman2004,Newman2016}, which compares the density of edges within the communities with the expected number if the vertices were attached at random, since a random graph is not expected to have a community structure.

   Despite the interdisciplinarity and the wide practical importance of finding communities for the study of e.g.\ metabolic process, marketing strategies and improving the routing in World Wide Web, community detection remains an open problem due to the high computational complexity of the optimization process required to uncover this structure in a network. However, many heuristic methods of modularity optimization show good agreement with peculiar relevant characteristics of real systems.  The aforementioned suggests that it could be advantageous to use the high effectiveness of SAWs in exploring the network structure for the implementation of a community detection algorithm.

   We use two properties of SAW for each par of vertices $i$ and $j$:\@ the probability for a walker departing from vertex $i$ to reach vertex $j$ before stopping and the average number of steps taken by walks that reach $j$.  As is shown below, the ratio of these properties allows us to identify some structural patterns related to the community structure. The algorithm proposed in this paper is based on this information and agglomerative hierarchical clustering \cite{Defays1977,Day1984}. We show that the algorithm matches or supplant the performance (with respect to modularity optimization) of the other traditional hierarchical methods used for comparison and, consequently, enhances the precision of community detection.

   The paper is organized as follows. In Sec.~\ref{sec:concepts_and_method}, we define some measurements related to SAW's dynamic and also describe the method to obtain information about the community structure using them. In Sec.~\ref{sec:results_and_discussion}, we present and discuss the results for two different synthetic network types and six well-known real networks. Finally, in Sec.~\ref{sec:conclusion} the paper closes showing the conclusions and perspectives for future works.

  \section{Basic concepts and Method}
  \label{sec:concepts_and_method}

    A self-avoiding random walk (SAW) is an alternative method to a basic walk where each vertex in the network is not revisited during the same walk. The walker always finds a possible path through the unvisited vertices and stop when no more paths are available. Due to this, the size of each path is limited by the number of vertices in network, $N$.

    $M_{i}$ walkers are started from vertex $i$ and $m_{i,j}$ is the number of them who visited vertex $j$ before stopping. Also the number of steps required for each one to reach $j$, $l_{i,j}^{w_{i}^k}$ (where $w_{i}^k$ is the $k$-th walker that starts from $i$) is evaluated. Based on this information, we define the transition probability as a fraction of walkers beginning the SAW at vertex $i$ and reaching vertex $j$ before stopping,
    \begin{equation}
      p_{i,j} = \frac{m_{i,j}}{M_{i}},
      \label{eq:p}
    \end{equation}
    and its average length,
    \begin{equation}
      \langle l_{i,j} \rangle = \frac{1}{m_{i,j}}\sum_{w_{i}^k}l_{i,j}^{w_{i}^k},
    \end{equation}
    where the sum is done over all $w_{i}^k$ walkers that pass through $j$. Equation~\ref{eq:p} measures how ease it is for walkers to reach vertex $j$ starting from vertex $i$, independently of the topological distances, so if there are many ways to find $j$ or the connection between $i$ and $j$ is trivial, $p_{i,j}$ tends to be high. On the other hand, complementary to $p_{i,j}$, $\langle l_{i,j} \rangle$ reveals topological distances of network by the average number of steps for the pair $i,j$.

    We combine these two in a single measure defining $f_{i,j}$,
    \begin{equation}
      f_{i,j} \equiv \frac{p_{i,j}}{\langle l_{i,j} \rangle}.
      \label{eq:f}
    \end{equation}
    The reasoning behind this definition is that the inverse of the average length quantifies the ``closeness'' of two vertices, but even vertices that can be reached in a small number of steps could be difficult to reach because only few paths exist between them. Therefore, we multiply the inverse of the average length by the transition probability. Note that $0 \le f_{i,j} \le 1$ for $i\ne j$ and that $f_{i,j} \ne f_{j,i}$ in general.  If no walker starting from $i$ passes through $j$, we use $f_{i,j} = 0$.

\begin{figure}
      \centering
      \includegraphics[scale=0.37]{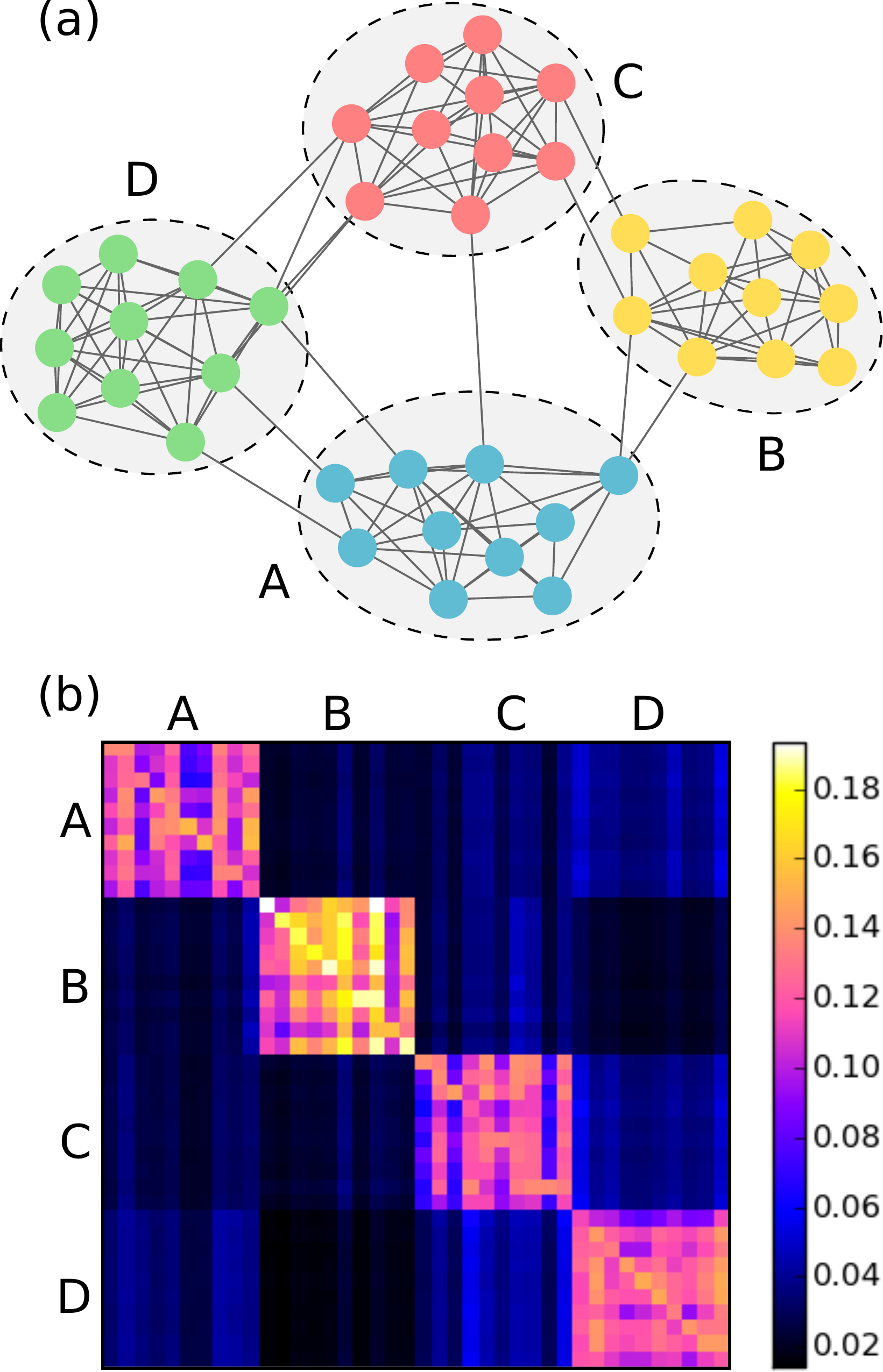}
      \caption{(a) A network displaying well-defined community structure with ten vertices in each one (vertices in the same community are sequentially numbered for easy visualization of the patterns in $f_{i,j}$). (b) Image representing the matrix of $f_{i,j}$ values, sub-matrices with high values appear for vertices in the same community. The diagonal values were filled according to equation~\ref{eq:f_ij}.}
      \label{fig:ex_rede}
    \end{figure}

    Networks which contain subgraphs of strongly connected vertices reveal some patterns in random walks that reflect high values of $f_{i,j}$ between elements in subgraphs and low values with the rest of the network. This behavior is related to communities. Although there is no precise definition of communities, they represent the concept of groups of vertices which are more densely connected inside the group than with the rest of the network \cite{Zhou2003,Newman:2006,Fortunato2010,W2016}. In fact, a set of vertices of the same community will tend to have relatively high values of $p_{i,j}$ and low values $\langle l_{i,j} \rangle$ (many walkers reach the members of the community within a few steps) due to the larger density of intra-community edges. Moreover, the perspective of the network by the walkers starting from different vertices in the same community tend to be similar (larger values of $f_{i,j}$ for vertices in the community, smaller values for the other vertices); this can be seen in Fig.~\ref{fig:ex_rede}. In this example, we have four well-defined communities with ten vertices in each one, sequentially numbered to favor the visualization of the matrix of $f_{i,j}$ in Fig.~\ref{fig:ex_rede}(b). It is easy to see that the regions with large values for $f_{i,j}$ correspond to the vertices in the same partition, since the walkers have similar perspectives of the network. Mathematically, the communities can be found by searching for groups of vertices whose lines in the matrix are similar.

    When constructing the matrix with the values of $f_{i,j}$ we are confronted with a difficulty: The value of $f_{i,i}$ is not defined, as the denominator in equation~\ref{eq:f} is zero.  Because our method described below depends on this matrix being defined, we use equation~\ref{eq:f} for $i\ne j$ and define
    \begin{equation}
      f_{i,i} = \max\limits_{\stackrel{1 \leq j \leq N}{j\ne i}} f_{i,j},
      \label{eq:f_ij}
    \end{equation}
    that is, we attribute to $f_{i,i}$ the same value of the vertex $j$ that is easiest to reach from $i$.  Our experiments show that this definition is appropriate for our method.  For the whole matrix we therefore have $0 \leq f_{i,j} \leq \max\limits_{i} f_{i,i}$.  The diagonal elements in the image of figure~\ref{fig:ex_rede} were filled according to equation \ref{eq:f_ij}.

    Given the matrix $\mathbf{F}$ of $f_{i,j}$ values as described above, we consider line $i$ as a vector of features associated with vertex $i$.  In order to obtain the best community distribution, we use principal component analysis (PCA) \cite{Abdi2010} to extract relevant information from the matrix, removing possible redundancies and trying to avoid the ``curse of dimensionality'', resulting in sparsely distributed data, impairing classification performance.  PCA is the standard procedure to reduce the dimensionality without loss of information \cite{Bellman1957} and the linear transformation proposed by it is
    \begin{equation}
      \widetilde{\mathbf{F}} = \mathbf{P} \cdot \mathbf{F},
    \end{equation}
    where $\mathbf{P}$ is the new basis composed by the principal components of $\mathbf{F}$ and $\widetilde{\mathbf{F}}$ is the data projected in $\mathbf{P}$ where the most relevant components (eigenvectors associated with higher eigenvalues) are chosen to classify the vertices.

    To compare two vertices $i$ and $j$ using $n$ principal components, we calculate the Bray-Curtis dissimilarity \cite{Kindt2005}
    \begin{equation}
      d(i,j) = \frac{\displaystyle\sum_{k=1}^{n}| \widetilde{f}(i,k) - \widetilde{f}(j,k)|}{\displaystyle\sum_{k=1}^{n} |\widetilde{f}(i,k) + \widetilde{f}(j,k)|}.
    \end{equation}
    If both vertices $i$ and $j$ belong to the same community, the perspective of the network from them is similar ($\widetilde{f}(i,k) \approx \widetilde{f}(j,k)$) and consequently $d(i,j)$ tend to be small.  We choose the average linkage method to merge two communities in an agglomerative method \cite{Day1984}. Starting with each vertex in its own community, at each step, the two groups with lowest Bray-Curtis dissimilarity join into a new one and create the new level of a dendrogram. The hierarchical clustering ends when all vertices compose a single community.

    To evaluate the partitioning, a frequently used measure is the modularity ($Q$) \cite{Newman2004,Newman06062006,Newman2016}. If $e_{ii}$ is the fraction of edges within group $i$ and $a_{i}$ is the fraction of edges connected to the vertices in community $i$, the modularity can be written as
    \begin{equation}
      Q = \sum_{i=1}^{n_{c}} (e_{ii} - a_{i}^{2}),
    \end{equation}
    where $n_{c}$ is the number of communities. When the number of within-community edges is the same as expected for random connections among the vertices we have $Q = 0$. On the other hand, values greater than zero indicate the presence of modular structures. Therefore, by this definition, the best community distribution is the one with the highest value of $Q$. Although intuitive, this measure has a resolution limitation that affect large networks and would fail to find small communities \cite{Fortunato2011,Fortunato2012,Fortunato2016}.

    The algorithm proposed for finding communities through SAW as follows.
    \begin{enumerate}
      \item Perform a sufficient number of SAWs starting from each vertex.
      \item Calculate the matrix $\mathbf{F}$ using equations~(\ref{eq:f}) and~(\ref{eq:f_ij}).
      \item Apply the PCA method to $\mathbf{F}$.
      \item Use two principal components of $\mathbf{F}$ to determine $\widetilde{\mathbf{F}}$.
      \item Find hierarchical clustering by average linkage and Bray-Curtis dissimilarity using the resulting $\widetilde{\mathbf{F}}$.
      \item Compute modularity for each level of the dendrogram.
      \item Save the configuration with maximum modularity.
      \item Increase the number of principal components and determine the corresponding new $\widetilde{\mathbf{F}}$.
      \item Repeat from step 5 until all PCA components are used.
      \item Choose the number of PCA components which results in the largest value of modularity.
    \end{enumerate}

    The time complexity of this algorithm is not so easy to determine, since the number of steps in SAW dynamics is not deterministic, but in the worst case we estimate the computation of $\mathbf{F}$ as taking $\mathcal{O}(MN^{2})$, where $M$ is the number of walker per vertex with at most $N$ hops each. The PCA method is $\mathcal{O}(N^{3})$ due to the time to calculate the covariance matrix and its eigenvalue decomposition. After that, hierarchical clustering and modularity are computed at $\mathcal{O}(N^{2}\log N)$ each one, for real graphs, and this must be repeated $N$ times (one for each number of principal components). Therefore, the time complexity of this technique is, in worst case $\mathcal{O}(MN^{2} + N^{3}\log N)$ but $M$ can be rewritten as $\alpha N$ with $\alpha$ a positive constant, so we finally find $\mathcal{O}(N^{3}\log N)$. For best time performance, our method can be parameterized through a fixed
    number of principal components and a reduced total number of walkers in the network, but this time savings could possibly result in a drop in precision.

  \section{Results and Discussion}
  \label{sec:results_and_discussion}

    We tested the performance of the community detection by SAW in artificial and real-world networks. For artificial networks, we used the traditional benchmark proposed by Girvan and Newman (GN)~\cite{Girvan2002} and the more flexible benchmark developed by Lancichinetti, Fortunato and Radicchi (LFR)~\cite{Lancichinetti2008}. For real cases, we applied our algorithm to six well-known networks, Zachary's karate club~\cite{Zachary1977}, bottlenose dolphins~\cite{Lusseau2003}, \textit{Les Mis\'{e}rables}~\cite{knuth1993}, American college football teams~\cite{Girvan2002}, jazz musicians~\cite{Gleiser2003} and \textit{C. elegans}~\cite{White1986,WattsDJ1998}. Although the sizes of these real networks are small, the advantage of their use is to allow us to check the result beyond modularity score, increasing trust in the partitions found.

    To validate the accuracy of our method we compare the results with four popular community detection algorithm. The first of them is the technique proposed by Girvan-Newman (GN)~\cite{Girvan2002}. The basic idea of this method is to identify inter-community links through edge betweenness centrality. Edges with the highest betweenness are removed from the graph step-by-step to obtain the community structure by hierarchical clustering. Although this method provides decompositions with high $Q$ values, it is quite slow because the edge betweenness centrality must be recalculated for all links in every iteration, rendering the method less useful for large networks. Due to this, Newman and Clauset developed an agglomerative hierarchical clustering algorithm based on a fast greedy technique (FG) that optimize modularity as vertices are joined into clusters~\cite{Clauset2004}. However, this implementation tends to form big communities and consequently decrease $Q$ score. The third algorithm used was developed by Pons and Latapy~\cite{Pons2005}, and utilizes traditional random walks to define a similarity measure between vertices and make an agglomerative hierarchical structure based on Ward's method. This approach is known as Walktrap (Wt) and (similar to FG), this algorithm, although fast, does not show high values of modularity for real networks. The last approach, known as Infomap (IMap), was propose by Rosvall and Bergstrom \cite{Rosvall2008} and as well as Wt, it also uses random walks dynamics to detect partitions. In this case, the algorithm compresses a description of information flow on the network resulting in a simplification of the graph that highlights community structures. In contrast to the three previous methods, IMap does not optimize the modularity function.

    In all these networks, we used $M_{i} = M = 10,000$, amounting to a total of $MN$ walkers, to ensure the stability of both measures $p_{i,j}$ and $\langle l_{i,j} \rangle$. In all networks tested $N \ll M$.

    \subsection{Benchmarks}

\begin{figure}
	\centering
	\includegraphics[scale=0.58]{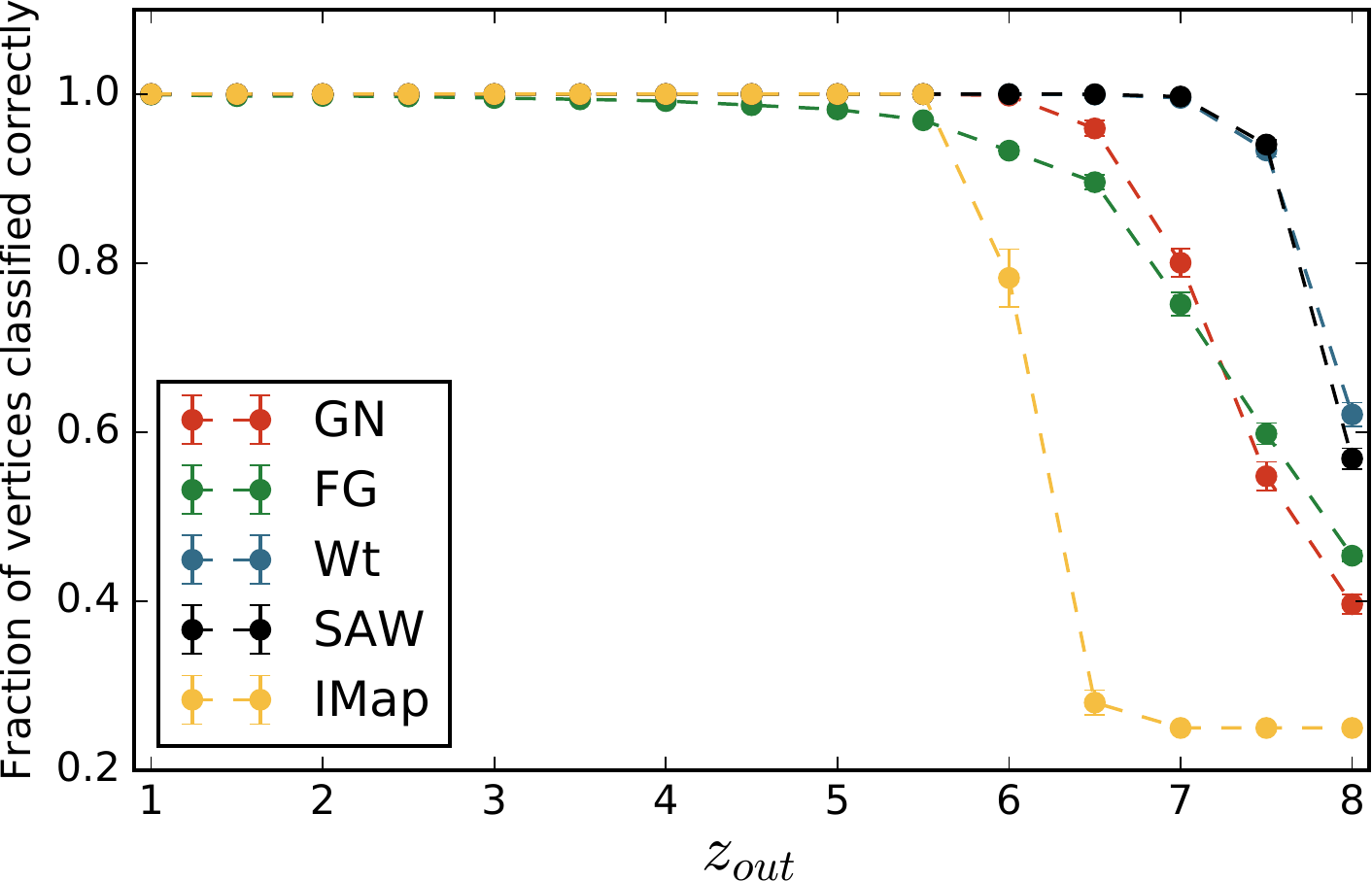}
	\caption{Fraction of vertices classified correctly by different partition algorithms tested on Girvan-Newnam benchmark. Each point is an average over 100 graph realization and the error bars correspond to one standard deviation. The Wt and SAW methods show the best results, since they lose the precision only at about $z_{out} = 7.5$. On the other hand IMap, the only algorithm that does not minimize the modularity, started decreasing quickly at $z_{out} = 6$.}
	\label{fig:gn_bench}
    \end{figure}

    The GN networks consist of 128 vertices split into four groups with 32 members, and average degree equal to 16. In this method, vertices belonging to the same group are randomly connected as in the Erd\H{o}s-R\'{e}nyi model with average in-group degree of $z_{in}$ and vertices of different groups are connected with average out-group degree $z_{out}$ such that $z_{in} + z_{out}=16$. In Fig.~\ref{fig:gn_bench} we show the efficiency of different methods as $z_{out}$ increases (i.e.\ the community structure becomes less well-defined). FG is the first algorithm to lose precision, around $z_{out} \sim 4$, followed by IMap that begins to fail at $z_{out} = 6$, and from this point, the quality of classification drops quickly until $z_{out} = 8$, a situation in which the internal degree is equal to the external degree. GN, in turn, begin to lose considerable accuracy for $z_{out} > 6$. Both Wt and the method proposed here have more accuracy in classification than the others, since the fraction of vertices classified correctly starts to decrease only near $z_{out} = 7.5$.

\begin{figure*}
	  \centering
	  \includegraphics[scale=0.35]{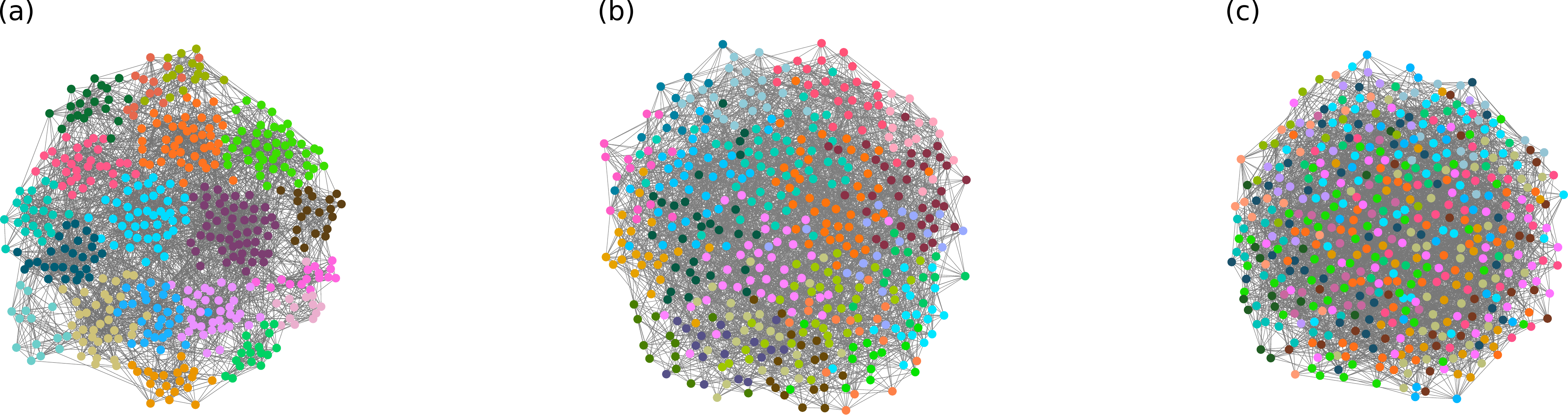}
	  \caption{Example of LFR benchmarks for $N = 500$, $\beta = 1$ and $\gamma = 3$ where the colors represents the distinct partitions. In (a) we have a graph with $18$ well defined communities generated with $\mu = 0.3$; in (b) for $\mu = 0.5$, the $21$ is is already hard to distinguish the partitions visually; finally in (c) the colors of the $17$ groups in the network generated with $\mu = 0.7$ are completely mixed, displaying the lack of a any definite community structure.}
	  \label{fig:lfr_ex}
      \end{figure*}

    The LFR benchmark is a synthetic network, which reproduces some features of realistic networks with a power law behavior of the degree distribution and the community sizes with exponents $\gamma$ and $\beta$, respectively. In this situation, each vertex shares a fraction of links $\mu$ with vertices in different partitions. This value is known as mixing parameter and changes in range $0 \le \mu \le 1$. When $\mu = 0.5$, the number of links in and out of communities is the same and it is difficult to clearly distinguish the partitions as shown in Fig. \ref{fig:lfr_ex}. In this example, we have three networks with $N = 500$ where the partitions are represented by colors and defined with $\beta = 1$, $\gamma = 3$ and $\mu = 0.3, 0.5, 0.7$ respectively. It is easy to see that as $\mu$ increases the communities become more difficult to distinguish. To quantify the precision of the divisions, this benchmark uses the normalized mutual information (NMI), a measure often used in tests of graph clustering algorithms \cite{Lancichinetti2009}. This measure equals one if all partitions match with the original division and zero if all vertices were classified incorrectly.

\begin{figure}
	  \centering
	  \includegraphics[scale=0.5]{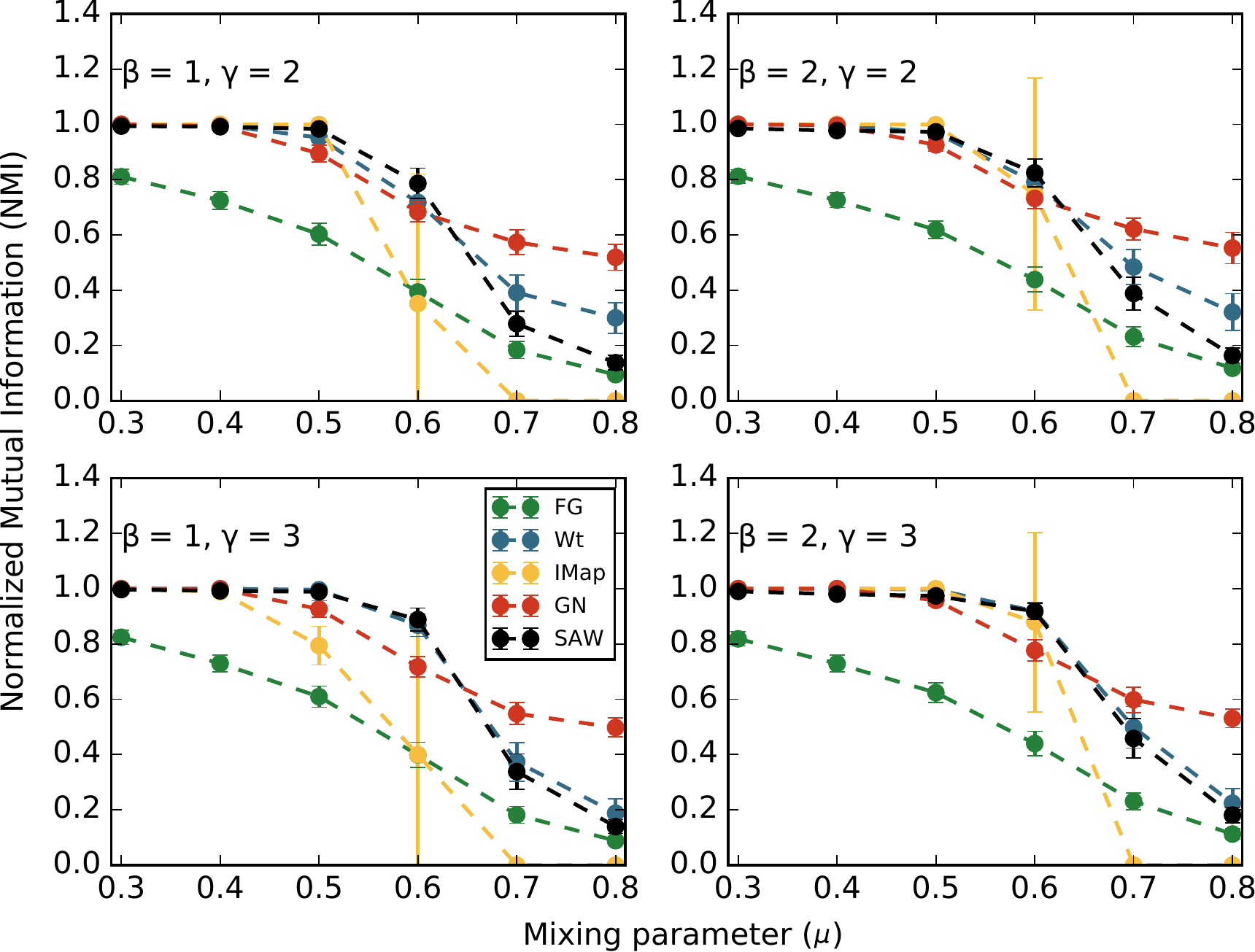}
	  \caption{Comparison of five different community detection algorithms on the LFR benchmark for distinct values of $\beta$ and $\gamma$. The parameters used were $N = 500$, $\langle k \rangle = 16$. The community sizes changed from $10$ to $50$ for each one of $100$ graph realization over $\mu$ with their respective standard deviation. The SAW algorithm produced good results for this benchmark.}
	  \label{fig:LFR}
      \end{figure}

      Fig.~\ref{fig:LFR} shows the comparison of the five different methods applied to the LFR benchmark with $N = 500$, $\langle k \rangle = 16$ and community sizes from $10$ to $50$. For the exponents we have chosen typical values of real networks in the ranges $1 \le \beta \le 2$ and $2 \le \gamma \le 3$. As for the GN benchmark, the FG algorithm, although fast, produced the worst results in all combinations of parameters $\beta$ and $\gamma$. While in the GN benchmark this technique shows good accuracy when partitions are well defined, in the LFR benchmark the NMI starts at a considerably lower value ($0.8$) when compared to the other methods, indicating that the method loses classification quality when applyed to more heterogeneous networks. IMap is the most sensitive method to variations of the parameter $\beta$. For $\beta = 1$, the NMI begins to decrease at $\mu = 0.5$ and goes to zero when $\mu > 0.6$. For $\beta = 2$, the drop in accuracy starts shortly after $\mu = 0.6$, together with the other algorithms, and goes to zero. The GN method shows good results for this benchmark, in all cases it has the smallest loss of NMI, begining at $\mu = 0.5$. The Wt and SAW methods have again similar behavior with good accuracy in all situation (specially for $\gamma = 3$), having a significant loss of accuracy after $\mu = 0.6$. We can thus conclude that community detection through SAW produces good results on benchmark networks, similar to or even surpassing the other four studied techniques, depending on network characteristics.

    \subsection{Real Networks}

    \begin{figure}
	\centering
	\includegraphics[scale=0.48]{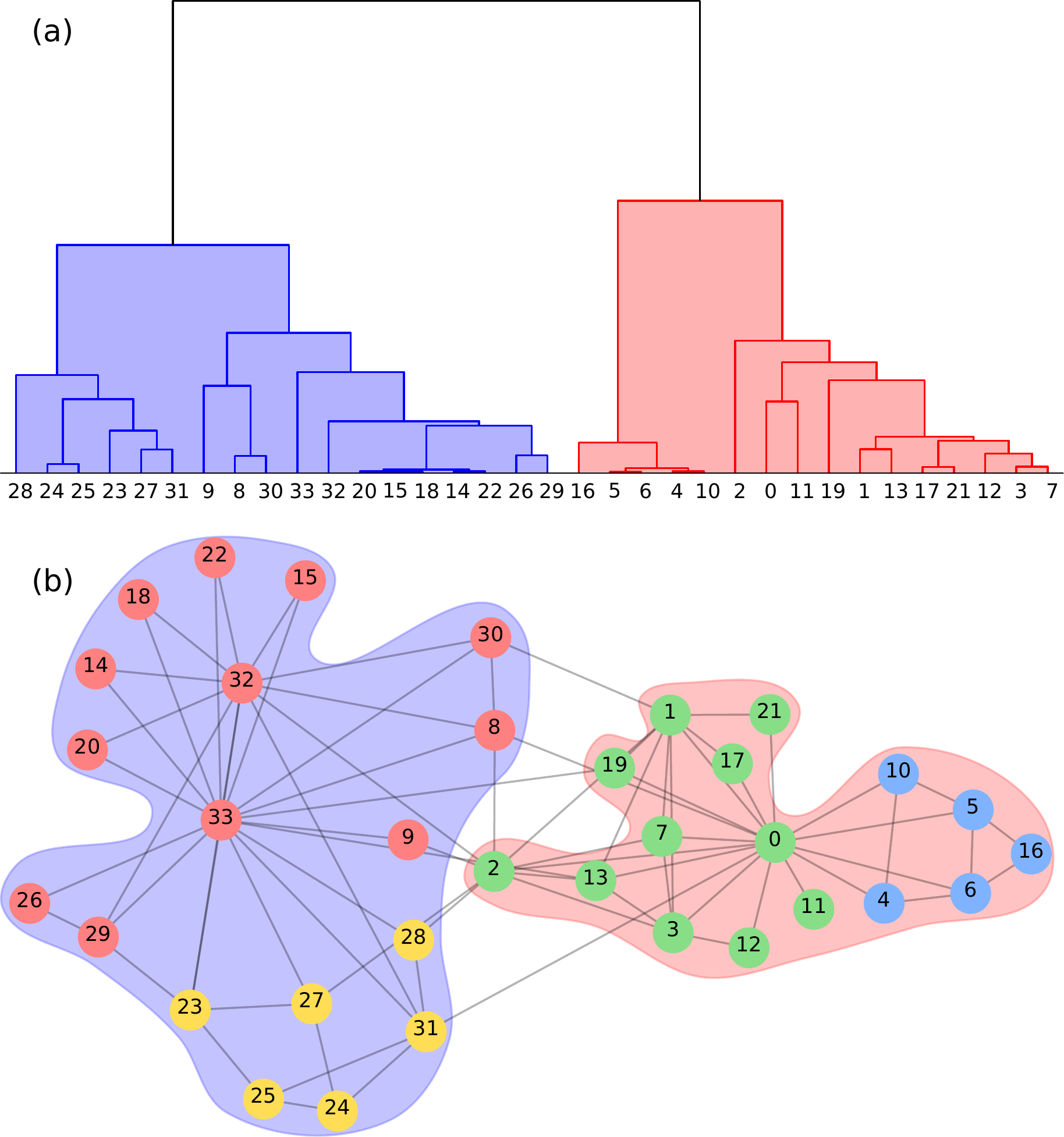}
	\caption{(a) Dendrogram obtained by our method for the Zachary's karate club network. The colors blue and red represent the club after split (except vertex 2). (b) Community structure which maximize the modularity function ($Q = 0.4197$). The blue and red areas follow the color code of the  dendrogram.}\label{fig:karate}
    \end{figure}

    \begin{figure*}
	\centering
	\includegraphics[scale=0.4]{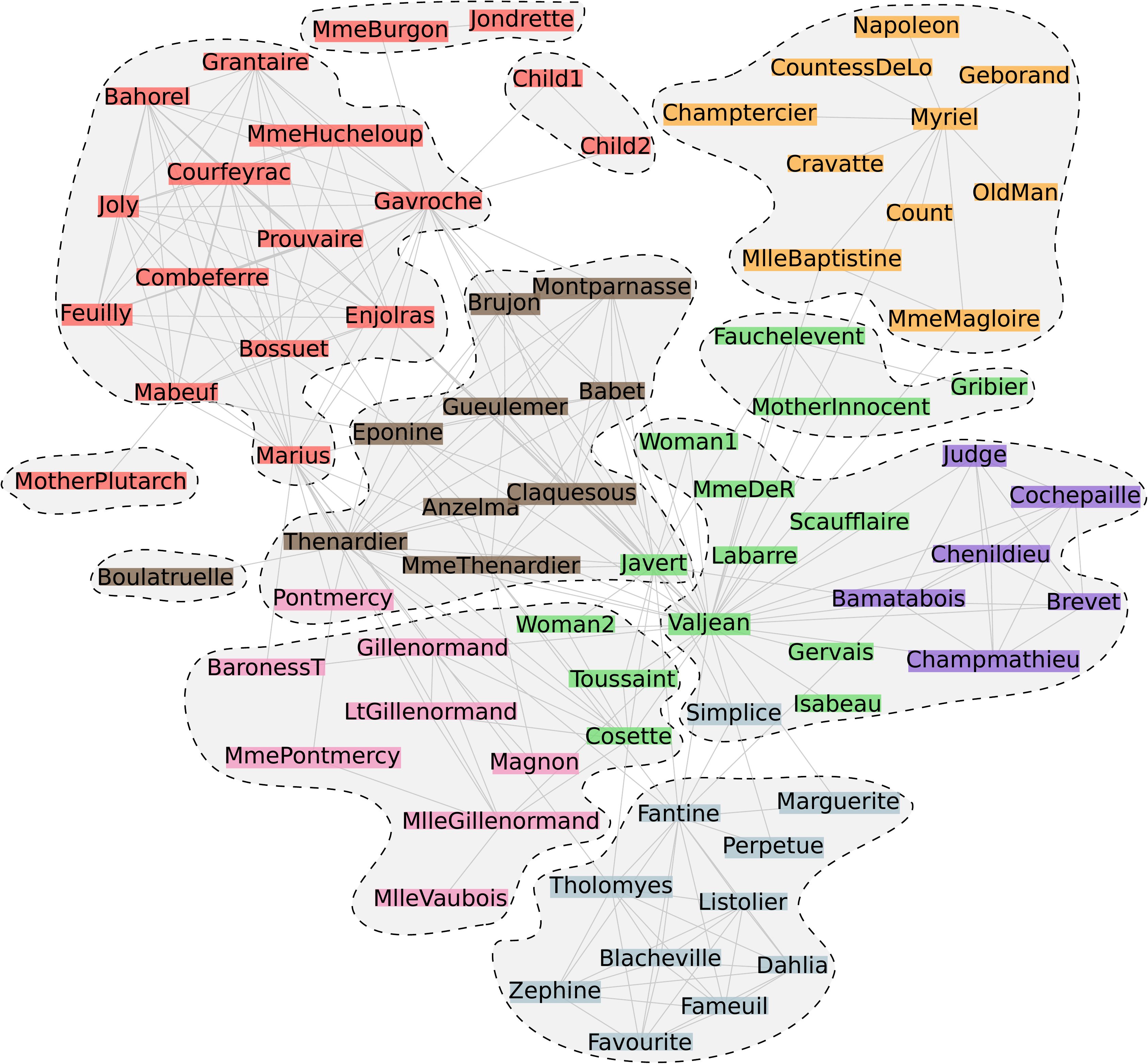}
	\caption{Community distribution in \textit{Les Miserables} network. The colors show the best divisions found by our algorithm which reached $Q = 0.5467$ with $7$ clusters. Areas limited by dashed lines are the distribution resulting from the GN method and represent $11$ groups with $Q = 0.5380$.}
      \label{fig:lesmis}
    \end{figure*}

    \begin{figure*}
	\centering
	\includegraphics[scale=0.7]{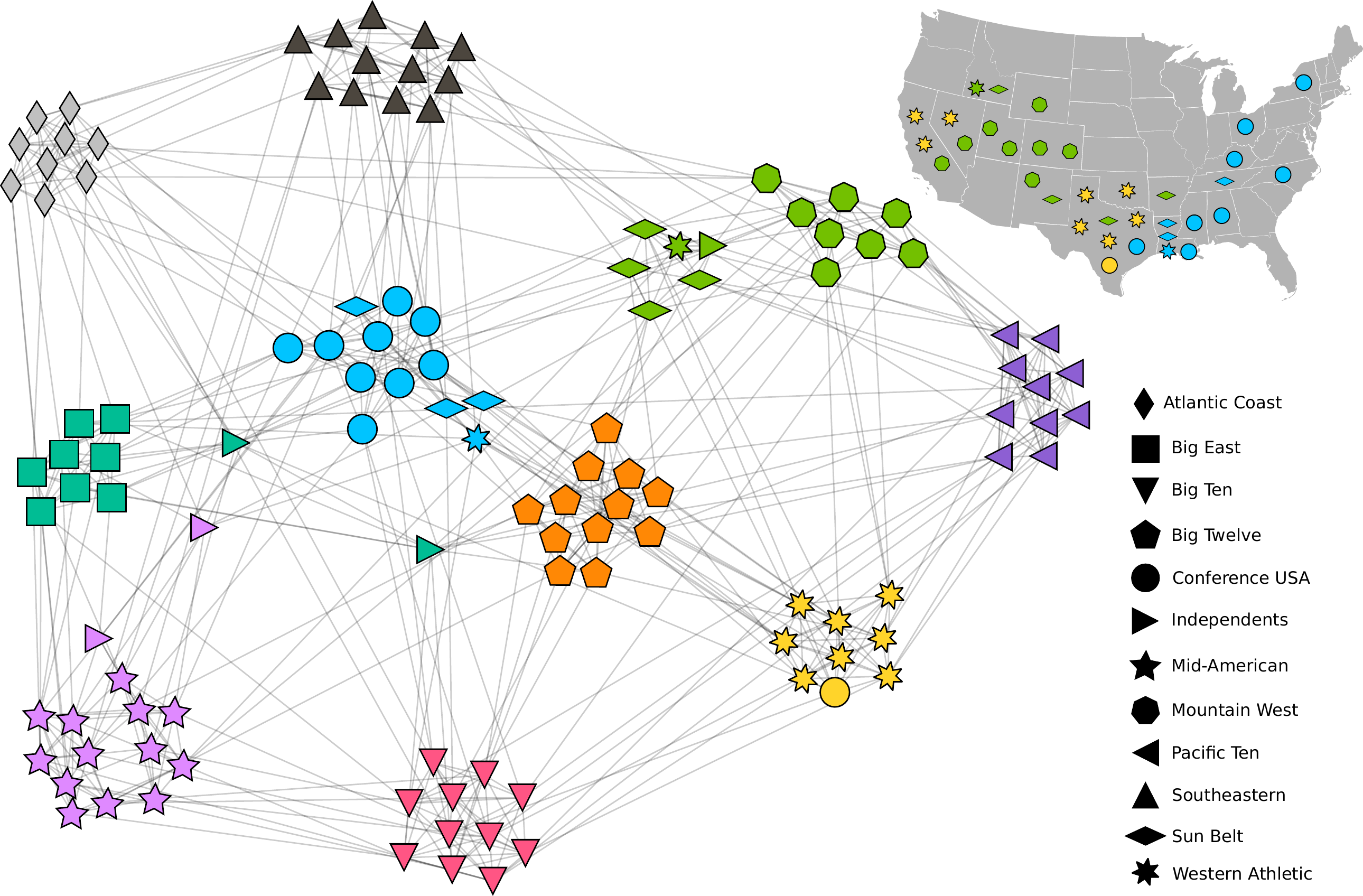}
	\caption{A network of American college football teams for the 2000 season. The vertices are colored according to community detection by SAW and geometric shapes represent the respective conference of the teams. The USA map indicates the influence of geographic localization in our classification (colleges geographically close play more with each other). This community distribution reaches $Q = 0.6044$ with $10$ divisions.}
      \label{fig:football}
    \end{figure*}

    Now we turn our attention to the real world networks. The first case is a social network presented by Wayne Zachary after he studied the friendship relations between $34$ members of a karate club at an American university. After three years of observation, $78$ links indicate the interaction of the participants inside and outside of club activities. At some point there was a disagreement between the administrator and the main teacher, and the result was the split into two smaller clubs. The dendrogram of the hierarchical structure found by our method for Zachary's network is shown in Fig.~\ref{fig:karate}(a) and we identify easily two big groups (red and blue regions) which represent the club after the split, except for vertex~2 (usually misclassified). In Fig.~\ref{fig:karate}(b) we show the community distribution which maximizes the modularity function, $Q = 0.4197$. This score was obtained with four communities that are subdivisions of the real division (represented by the blue and red areas) and it is the highest value ever found in literature through a variety of heuristic techniques \cite{Donetti2004,Hua-Wei2009,Zhang2012}. To reach this configuration with our method, the three principal components of the PCA were used.

    Fig.~\ref{fig:lesmis} shows the network of interactions between the characters in the novel \textit{Les Miserables}, written by Victor Hugo in the 19th century.  In this system the $77$ vertices represent the actors and the $254$ links indicate co-appearance in one or more scenes. The colors show the community structure found by the SAW method ($Q = 0.5467$) and the areas bounded by dash lines are the subdivisions suggested by GN algorithm ($Q = 0.5380$). Although these detections result in similar modularity scores, the partitions are completely different. In this example, we notice that partitions with similar values of modularity can be very different.

    The split into $7$ clusters reveals a variety of social interactions along the novel. The main community concentrates three major characters, Jean Valjean, Javert and Cosette. Fantine, Cosette's mother, is the center of the  group composed by her friends and her husband Tholomyes. The other important subplot is the partition which reflects the Friends of the ABC. Marius's family stays in a particular community as well as Cosette's foster family. The bishop Myriel is the main representative in orange group and the smaller division (purple) has a judge and outlaw people. To achieve this community distribution our algorithm uses the information of seven principal components.

    Observing the two results shown in Fig~\ref{fig:lesmis}, we find three main differences. First, communities with few members (in the GN decomposition) have been joined in a big one (in the SAW decomposition). Second, in the GN case, Valjean, Javert and Cosette belong to distinct groups, while for the SAW method they are in the same. This happens because these vertices are hubs and the walkers starting from one of them quickly reach the others, so the perspective of the network is very similar for them. Finally, we have some individual characters which were classified in a different way as Sister Simplice: she has four edges with characters in two different communities, so in this situation, both classifications make sense.

    The last example of applying our method to real networks is shown in Fig.~\ref{fig:football}, a network of Division~I American college football teams for the $2000$ season. The $115$~vertices represent the teams and a game between two of them corresponds to a link. In this competition, the teams were organized in $12$ conferences, games intra-conference being more frequent than inter-conference, so a community structure arises from these organizations. Nevertheless inter-conference matches are not uniformly divided, the geographic position of teams also influence in the games (colleges geographically close play more with each other).

    The colors represent the community structure found by the SAW method and the shapes correspond to the conferences. We were able to classify perfectly all teams affiliated in five conferences. The Sun Belt's teams have been split into two communities, due to geographic localization, as showed on the USA map, where teams with the same color are geographically close despite belonging to different conferences. These effects can also be seen in Western Athletic conference. Another interesting fact happened with independent teams: as they do not belong to any conference, the algorithm tends to group them with those more closely associated. In fact, in $2002$, the UCF Knights joined the Mid-American conference and in $2003$ the Utah State Aggies joined the Sun Belt conference. This distribution with $10$ communities reached a high modularity score $Q = 0.6044$, using eleven principal components.

    \begin{table*}[]
      \centering
      \caption{Comparison of the modularity and number of groups (inside parenthesis) for different community detection algorithms applied in some real-world networks.}
      \label{tab:modularity}
      \begin{ruledtabular}
	\begin{tabular}{lcccccc}
	  Network    & $N$ &     FG     &     GN     &     Wt     & IMap 	&     SAW      \\ \hline
	  Karate     & 34  & 0.3807 (3) & 0.4013 (5) & 0.3532 (5) & 0.4020 (3)	&   0.4197 (4) \\
	  Dolphins   & 62  & 0.4955 (4) & 0.5194 (5) & 0.4888 (4) & 0.5247 (6) 	&   0.5277 (5) \\
	  Lesmis     & 77  & 0.5006 (5) & 0.5380 (11)& 0.5214 (8) & 0.5461 (9)	&   0.5467 (7) \\
	  Football   & 115 & 0.5497 (6) & 0.5996 (10)& 0.6029 (10)& 0.6005 (12)	&   0.6044 (10)\\
	  Jazz       & 198 & 0.4389 (4) & 0.4051 (39)& 0.4384 (11)& 0.2800 (7)	&   0.4428 (4) \\
	  C. elegans & 297 & 0.3723 (5) & 0.3018 (33)& 0.3532 (22)& 0.3858 (8)	&   0.3746 (5) \\
	\end{tabular}
      \end{ruledtabular}
    \end{table*}

    Table~\ref{tab:modularity} shows the performance of community detection through SAW in comparison with other popular methods applied to the some real-world networks. Besides Zachary's karate, \textit{Les Miserables} (Lesmis) and American college football, we also used a social network of $62$ bottlenose dolphins studied in New Zealand, a jazz musicians network, where there is a link between two of $198$ musicians if they have played together in a band, and the neural network of the worm \textit{Caenorhabditis  elegans}. Note that
    the GN and FG methods have no stochastic elements, and always generate the
    same results for the same network. This is also valid for the Wt method,
    because although it is based on random walks, only their steady state as captured by the transition probabilities (computed from the adjacency matrix)
    is used.  In the method proposed here (SAW), the stochastic character of the random walks does not influence the results, as a large enough number of walkers is computed to assure that we are close to the steady state value of the $f_{i,j}$ matrix elements, and the method is deterministic given this matrix.  The only method that presents some variations from one execution to another is Infomap. In this case, we executed it 50 times for each network, seeing no variations (to the fifth decimal place) for the karate, football and jazz networks; for the other networks, the standard deviations were: 0.0036 for the dolphins, 0.0010 for Lesmis, and 0.0150 for C. Elegans.

    With the exception of C. elegans, in all other cases our method presented the highest modularity, though in some situations the $Q$ score of another algorithm was very close. However, this small difference is enough to produce considerably distinct results, e.g., in the jazz network, where modularity for FG and Wt differ from SAW only to $\Delta Q \sim 0.004$  eleven groups are found by Wt while FG and SAW split the network in four parts. The same situation also happens in the Lesmis network with GN and SAW as discussed in Fig. \ref{fig:lesmis}. On the other hand, FG and SAW produces modularity quite similar ($Q = 0.3723$ and $Q = 0.3746$, respectively) in C.\ Elegans network but IMap shows a better modularity score with eight partitions. It is interesting to note how the Wt method, which achieves good results for synthetic networks, underperformed in many of the real networks.

    The results in Table~\ref{tab:modularity} suggest that the method here proposed is able to determine a good community distribution. But it must be mentioned that the restricted walker can be computationally expensive. So our new dynamic approach to find the best network division is indicated for not so large networks.

  \section{Conclusion}
  \label{sec:conclusion}

    Earlier works revealed that self-avoiding random walks are very efficient to search and explore networks, suggesting they could be a useful mechanism to find modular structures. In this paper, we proposed a community detection method which quantifies the similarity between vertices based on this restrict random walk.

    Experiments with the GN and LFR benchmarks show that our method produces good results for different parameter combinations as network size, degree distribution and number of communities. We used these tests also to compare performance with other four popular community detection algorithms (GN, FG, Wt and IMap), and we found that this new technique provides classifications of similar (or better) quality as usual approaches for synthetic networks. We also applied it on six well-known real-world cases and found higher modularity scores than traditional methods, revealing a better community distribution with great agreement with reality.  It is an interesting question for further study to find why this happens, as it can point to better benchmark models for community detection.  The possibility is still open that the precision of the method could be increased by the use of other similarity measures or classification methods, from which many are available in the literature.

    The method has a high computational cost due to the need of computing a large number of random walks, as SAWs are not a Markovian stochastic process, however it achieves high modularity scores with good agreement in artificial and real-world networks.  Therefore it is recommended to study ways of speeding its execution up to be able to use it for large networks.

    It is straightforward to generalized the algorithm for weighted networks (following edges with probability proportional to their weight) and to directed networks (following edges only in the their directions, although care must be taken with unreachable components) as well as to determine communities of individual vertices with local exploration by SAWs (the precise details will depend on a definition for the local community of a vertex). Another possibility is to adapt the method to better understand the impact of the metadata in community structure (using metadata information in the random walk decisions). Finally, other possibilities to study the network structures through the matrix $\mathbf{F}$ (beside communities) are also open.

  \section{Acknowledgments}

    The authors are grateful to Capes for financial support and to the Information Technology Superintendence (STI) of the University of S\~ao Paulo for providing HPC resources.

  \bibliographystyle{apsrev4-1}
  \bibliography{mybibliography}

\begin{thebibliography}{59}%
\makeatletter
\providecommand \@ifxundefined [1]{%
 \@ifx{#1\undefined}
}%
\providecommand \@ifnum [1]{%
 \ifnum #1\expandafter \@firstoftwo
 \else \expandafter \@secondoftwo
 \fi
}%
\providecommand \@ifx [1]{%
 \ifx #1\expandafter \@firstoftwo
 \else \expandafter \@secondoftwo
 \fi
}%
\providecommand \natexlab [1]{#1}%
\providecommand \enquote  [1]{``#1''}%
\providecommand \bibnamefont  [1]{#1}%
\providecommand \bibfnamefont [1]{#1}%
\providecommand \citenamefont [1]{#1}%
\providecommand \href@noop [0]{\@secondoftwo}%
\providecommand \href [0]{\begingroup \@sanitize@url \@href}%
\providecommand \@href[1]{\@@startlink{#1}\@@href}%
\providecommand \@@href[1]{\endgroup#1\@@endlink}%
\providecommand \@sanitize@url [0]{\catcode `\\12\catcode `\$12\catcode
  `\&12\catcode `\#12\catcode `\^12\catcode `\_12\catcode `\%12\relax}%
\providecommand \@@startlink[1]{}%
\providecommand \@@endlink[0]{}%
\providecommand \url  [0]{\begingroup\@sanitize@url \@url }%
\providecommand \@url [1]{\endgroup\@href {#1}{\urlprefix }}%
\providecommand \urlprefix  [0]{URL }%
\providecommand \Eprint [0]{\href }%
\providecommand \doibase [0]{http://dx.doi.org/}%
\providecommand \selectlanguage [0]{\@gobble}%
\providecommand \bibinfo  [0]{\@secondoftwo}%
\providecommand \bibfield  [0]{\@secondoftwo}%
\providecommand \translation [1]{[#1]}%
\providecommand \BibitemOpen [0]{}%
\providecommand \bibitemStop [0]{}%
\providecommand \bibitemNoStop [0]{.\EOS\space}%
\providecommand \EOS [0]{\spacefactor3000\relax}%
\providecommand \BibitemShut  [1]{\csname bibitem#1\endcsname}%
\let\auto@bib@innerbib\@empty
%</preamble>
\bibitem [{\citenamefont {Stauffer}\ and\ \citenamefont
  {Aharony}(1971)}]{Stauffer1971}%
  \BibitemOpen
  \bibfield  {author} {\bibinfo {author} {\bibfnamefont {D.}~\bibnamefont
  {Stauffer}}\ and\ \bibinfo {author} {\bibfnamefont {A.}~\bibnamefont
  {Aharony}},\ }\href@noop {} {\emph {\bibinfo {title} {{Introduction to
  Percolation Theory}}}}\ (\bibinfo  {publisher} {Oxford University Press, New
  York},\ \bibinfo {year} {1971})\BibitemShut {NoStop}%
\bibitem [{\citenamefont {Arenas}\ \emph {et~al.}(2006)\citenamefont {Arenas},
  \citenamefont {D\'{\i}az-Guilera},\ and\ \citenamefont
  {P\'erez-Vicente}}]{Arenas2006}%
  \BibitemOpen
  \bibfield  {author} {\bibinfo {author} {\bibfnamefont {A.}~\bibnamefont
  {Arenas}}, \bibinfo {author} {\bibfnamefont {A.}~\bibnamefont
  {D\'{\i}az-Guilera}}, \ and\ \bibinfo {author} {\bibfnamefont {C.~J.}\
  \bibnamefont {P\'erez-Vicente}},\ }\href@noop {} {\bibfield  {journal}
  {\bibinfo  {journal} {Phys. Rev. Lett.}\ }\textbf {\bibinfo {volume} {96}},\
  \bibinfo {pages} {114102} (\bibinfo {year} {2006})}\BibitemShut {NoStop}%
\bibitem [{\citenamefont {Pastor-Satorras}\ and\ \citenamefont
  {Vespignani}(2001)}]{Pastor-Satorras2001}%
  \BibitemOpen
  \bibfield  {author} {\bibinfo {author} {\bibfnamefont {R.}~\bibnamefont
  {Pastor-Satorras}}\ and\ \bibinfo {author} {\bibfnamefont {A.}~\bibnamefont
  {Vespignani}},\ }\href@noop {} {\bibfield  {journal} {\bibinfo  {journal}
  {Phys. Rev. Lett.}\ }\textbf {\bibinfo {volume} {86}},\ \bibinfo {pages}
  {3200} (\bibinfo {year} {2001})}\BibitemShut {NoStop}%
\bibitem [{\citenamefont {Lahtinen}\ \emph {et~al.}(2001)\citenamefont
  {Lahtinen}, \citenamefont {Kert{\'{e}}sz},\ and\ \citenamefont
  {Kaski}}]{Lahtinen2001}%
  \BibitemOpen
  \bibfield  {author} {\bibinfo {author} {\bibfnamefont {J.}~\bibnamefont
  {Lahtinen}}, \bibinfo {author} {\bibfnamefont {J.}~\bibnamefont
  {Kert{\'{e}}sz}}, \ and\ \bibinfo {author} {\bibfnamefont {K.}~\bibnamefont
  {Kaski}},\ }\href@noop {} {\bibfield  {journal} {\bibinfo  {journal}
  {Physical Review E}\ }\textbf {\bibinfo {volume} {64}},\ \bibinfo {pages}
  {057105} (\bibinfo {year} {2001})}\BibitemShut {NoStop}%
\bibitem [{\citenamefont {Almaas}\ \emph {et~al.}(2003)\citenamefont {Almaas},
  \citenamefont {Kulkarni},\ and\ \citenamefont {Stroud}}]{Almaas2003}%
  \BibitemOpen
  \bibfield  {author} {\bibinfo {author} {\bibfnamefont {E.}~\bibnamefont
  {Almaas}}, \bibinfo {author} {\bibfnamefont {R.}~\bibnamefont {Kulkarni}}, \
  and\ \bibinfo {author} {\bibfnamefont {D.}~\bibnamefont {Stroud}},\
  }\href@noop {} {\bibfield  {journal} {\bibinfo  {journal} {Physical Review
  E}\ }\textbf {\bibinfo {volume} {68}},\ \bibinfo {pages} {056105} (\bibinfo
  {year} {2003})}\BibitemShut {NoStop}%
\bibitem [{\citenamefont {Noh}(2004)}]{Noh2004}%
  \BibitemOpen
  \bibfield  {author} {\bibinfo {author} {\bibfnamefont {J.~D.}\ \bibnamefont
  {Noh}},\ }\href@noop {} {\bibfield  {journal} {\bibinfo  {journal} {Physical
  Review Letters}\ }\textbf {\bibinfo {volume} {92}},\ \bibinfo {pages}
  {118701} (\bibinfo {year} {2004})}\BibitemShut {NoStop}%
\bibitem [{\citenamefont {Masuda}\ and\ \citenamefont
  {Konno}(2004)}]{Masuda2004}%
  \BibitemOpen
  \bibfield  {author} {\bibinfo {author} {\bibfnamefont {N.}~\bibnamefont
  {Masuda}}\ and\ \bibinfo {author} {\bibfnamefont {N.}~\bibnamefont {Konno}},\
  }\href@noop {} {\bibfield  {journal} {\bibinfo  {journal} {Physical Review E
  - Statistical, Nonlinear, and Soft Matter Physics}\ }\textbf {\bibinfo
  {volume} {69}},\ \bibinfo {pages} {1} (\bibinfo {year} {2004})}\BibitemShut
  {NoStop}%
\bibitem [{\citenamefont {Tejedor}\ \emph {et~al.}(2009)\citenamefont
  {Tejedor}, \citenamefont {B{\'{e}}nichou},\ and\ \citenamefont
  {Voituriez}}]{Tejedor2009}%
  \BibitemOpen
  \bibfield  {author} {\bibinfo {author} {\bibfnamefont {V.}~\bibnamefont
  {Tejedor}}, \bibinfo {author} {\bibfnamefont {O.}~\bibnamefont
  {B{\'{e}}nichou}}, \ and\ \bibinfo {author} {\bibfnamefont {R.}~\bibnamefont
  {Voituriez}},\ }\href@noop {} {\bibfield  {journal} {\bibinfo  {journal}
  {Physical Review E}\ }\textbf {\bibinfo {volume} {80}},\ \bibinfo {pages}
  {065104} (\bibinfo {year} {2009})}\BibitemShut {NoStop}%
\bibitem [{\citenamefont {Tadi}(2001)}]{Tadi2001}%
  \BibitemOpen
  \bibfield  {author} {\bibinfo {author} {\bibfnamefont {B.}~\bibnamefont
  {Tadi}},\ }\href@noop {} {\bibfield  {journal} {\bibinfo  {journal} {The
  European Physical Journal B}\ }\textbf {\bibinfo {volume} {228}},\ \bibinfo
  {pages} {221} (\bibinfo {year} {2001})}\BibitemShut {NoStop}%
\bibitem [{\citenamefont {Comin}\ \emph {et~al.}(2014)\citenamefont {Comin},
  \citenamefont {Viana}, \citenamefont {Antiqueira},\ and\ \citenamefont
  {Costa}}]{Comin2014}%
  \BibitemOpen
  \bibfield  {author} {\bibinfo {author} {\bibfnamefont {C.~H.}\ \bibnamefont
  {Comin}}, \bibinfo {author} {\bibfnamefont {M.~P.}\ \bibnamefont {Viana}},
  \bibinfo {author} {\bibfnamefont {L.}~\bibnamefont {Antiqueira}}, \ and\
  \bibinfo {author} {\bibfnamefont {L.~D.~F.}\ \bibnamefont {Costa}},\
  }\href@noop {} {\bibfield  {journal} {\bibinfo  {journal} {Journal of
  Statistical Mechanics: Theory and Experiment}\ }\textbf {\bibinfo {volume}
  {2014}},\ \bibinfo {pages} {P12003} (\bibinfo {year} {2014})}\BibitemShut
  {NoStop}%
\bibitem [{\citenamefont {Aksoy}\ \emph {et~al.}(2016)\citenamefont {Aksoy},
  \citenamefont {Chung},\ and\ \citenamefont {Peng}}]{Aksoy2016}%
  \BibitemOpen
  \bibfield  {author} {\bibinfo {author} {\bibfnamefont {S.}~\bibnamefont
  {Aksoy}}, \bibinfo {author} {\bibfnamefont {F.}~\bibnamefont {Chung}}, \ and\
  \bibinfo {author} {\bibfnamefont {X.}~\bibnamefont {Peng}},\ }\href@noop {}
  {\bibfield  {journal} {\bibinfo  {journal} {ArXiv}\ } (\bibinfo {year}
  {2016})}\BibitemShut {NoStop}%
\bibitem [{\citenamefont {Lambiotte}\ \emph {et~al.}(2014)\citenamefont
  {Lambiotte}, \citenamefont {Salnikov},\ and\ \citenamefont
  {Rosvall}}]{Lambiotte2014}%
  \BibitemOpen
  \bibfield  {author} {\bibinfo {author} {\bibfnamefont {R.}~\bibnamefont
  {Lambiotte}}, \bibinfo {author} {\bibfnamefont {V.}~\bibnamefont {Salnikov}},
  \ and\ \bibinfo {author} {\bibfnamefont {M.}~\bibnamefont {Rosvall}},\
  }\href@noop {} {\bibfield  {journal} {\bibinfo  {journal} {Journal of Complex
  Networks}\ } (\bibinfo {year} {2014})}\BibitemShut {NoStop}%
\bibitem [{\citenamefont {Weng}\ \emph {et~al.}(2014)\citenamefont {Weng},
  \citenamefont {Zhao}, \citenamefont {Small},\ and\ \citenamefont
  {Huang}}]{Weng2014}%
  \BibitemOpen
  \bibfield  {author} {\bibinfo {author} {\bibfnamefont {T.}~\bibnamefont
  {Weng}}, \bibinfo {author} {\bibfnamefont {Y.}~\bibnamefont {Zhao}}, \bibinfo
  {author} {\bibfnamefont {M.}~\bibnamefont {Small}}, \ and\ \bibinfo {author}
  {\bibfnamefont {D.~D.}\ \bibnamefont {Huang}},\ }\href@noop {} {\bibfield
  {journal} {\bibinfo  {journal} {Physical Review E}\ }\textbf {\bibinfo
  {volume} {90}},\ \bibinfo {pages} {022804} (\bibinfo {year}
  {2014})}\BibitemShut {NoStop}%
\bibitem [{\citenamefont {Yang}(2005)}]{Yang2005}%
  \BibitemOpen
  \bibfield  {author} {\bibinfo {author} {\bibfnamefont {S.-J.}\ \bibnamefont
  {Yang}},\ }\href@noop {} {\bibfield  {journal} {\bibinfo  {journal} {Physical
  Review E}\ }\textbf {\bibinfo {volume} {71}},\ \bibinfo {pages} {016107}
  (\bibinfo {year} {2005})}\BibitemShut {NoStop}%
\bibitem [{\citenamefont {Costa}\ and\ \citenamefont
  {Travieso}(2007)}]{Costa2007}%
  \BibitemOpen
  \bibfield  {author} {\bibinfo {author} {\bibfnamefont {L.~D.~F.}\
  \bibnamefont {Costa}}\ and\ \bibinfo {author} {\bibfnamefont
  {G.}~\bibnamefont {Travieso}},\ }\href@noop {} {\bibfield  {journal}
  {\bibinfo  {journal} {Physical Review E}\ }\textbf {\bibinfo {volume} {75}},\
  \bibinfo {pages} {016102} (\bibinfo {year} {2007})}\BibitemShut {NoStop}%
\bibitem [{\citenamefont {{L{\'{o}}pez Mill{\'{a}}n}}\ \emph
  {et~al.}(2011)\citenamefont {{L{\'{o}}pez Mill{\'{a}}n}}, \citenamefont
  {Cholvi}, \citenamefont {L{\'{o}}pez},\ and\ \citenamefont {{Fern{\'{a}}ndez
  Anta}}}]{LopezMillan2011}%
  \BibitemOpen
  \bibfield  {author} {\bibinfo {author} {\bibfnamefont {V.~M.}\ \bibnamefont
  {{L{\'{o}}pez Mill{\'{a}}n}}}, \bibinfo {author} {\bibfnamefont
  {V.}~\bibnamefont {Cholvi}}, \bibinfo {author} {\bibfnamefont
  {L.}~\bibnamefont {L{\'{o}}pez}}, \ and\ \bibinfo {author} {\bibfnamefont
  {A.}~\bibnamefont {{Fern{\'{a}}ndez Anta}}},\ }\href {\doibase
  10.1002/net.20461} {\bibfield  {journal} {\bibinfo  {journal} {Networks}\ }
  (\bibinfo {year} {2011}),\ 10.1002/net.20461}\BibitemShut {NoStop}%
\bibitem [{\citenamefont {Marshak}\ \emph {et~al.}(2016)\citenamefont
  {Marshak}, \citenamefont {Rombach}, \citenamefont {Bertozzi},\ and\
  \citenamefont {D'Orsogna}}]{Marshak2016}%
  \BibitemOpen
  \bibfield  {author} {\bibinfo {author} {\bibfnamefont {C.~Z.}\ \bibnamefont
  {Marshak}}, \bibinfo {author} {\bibfnamefont {M.~P.}\ \bibnamefont
  {Rombach}}, \bibinfo {author} {\bibfnamefont {A.~L.}\ \bibnamefont
  {Bertozzi}}, \ and\ \bibinfo {author} {\bibfnamefont {M.~R.}\ \bibnamefont
  {D'Orsogna}},\ }\href@noop {} {\bibfield  {journal} {\bibinfo  {journal}
  {Physical Review E}\ }\textbf {\bibinfo {volume} {93}},\ \bibinfo {pages}
  {022308} (\bibinfo {year} {2016})}\BibitemShut {NoStop}%
\bibitem [{\citenamefont {Guo}\ \emph {et~al.}(2016)\citenamefont {Guo},
  \citenamefont {Cozzo}, \citenamefont {Zheng},\ and\ \citenamefont
  {Moreno}}]{Guo2016}%
  \BibitemOpen
  \bibfield  {author} {\bibinfo {author} {\bibfnamefont {Q.}~\bibnamefont
  {Guo}}, \bibinfo {author} {\bibfnamefont {E.}~\bibnamefont {Cozzo}}, \bibinfo
  {author} {\bibfnamefont {Z.}~\bibnamefont {Zheng}}, \ and\ \bibinfo {author}
  {\bibfnamefont {Y.}~\bibnamefont {Moreno}},\ }\href@noop {} {\bibfield
  {journal} {\bibinfo  {journal} {ArXiv}\ ,\ \bibinfo {pages} {32}} (\bibinfo
  {year} {2016})}\BibitemShut {NoStop}%
\bibitem [{\citenamefont {Kim}\ \emph {et~al.}(2016)\citenamefont {Kim},
  \citenamefont {Kyoung},\ and\ \citenamefont {Lee}}]{Kim2016}%
  \BibitemOpen
  \bibfield  {author} {\bibinfo {author} {\bibfnamefont {K.}~\bibnamefont
  {Kim}}, \bibinfo {author} {\bibfnamefont {J.}~\bibnamefont {Kyoung}}, \ and\
  \bibinfo {author} {\bibfnamefont {D.-S.}\ \bibnamefont {Lee}},\ }\href@noop
  {} {\bibfield  {journal} {\bibinfo  {journal} {Physical Review E}\ }\textbf
  {\bibinfo {volume} {93}},\ \bibinfo {pages} {052310} (\bibinfo {year}
  {2016})}\BibitemShut {NoStop}%
\bibitem [{\citenamefont {Herrero}\ and\ \citenamefont
  {Saboy{\'{a}}}(2003)}]{Herrero2003}%
  \BibitemOpen
  \bibfield  {author} {\bibinfo {author} {\bibfnamefont {C.~P.}\ \bibnamefont
  {Herrero}}\ and\ \bibinfo {author} {\bibfnamefont {M.}~\bibnamefont
  {Saboy{\'{a}}}},\ }\href@noop {} {\bibfield  {journal} {\bibinfo  {journal}
  {Physical Review E}\ }\textbf {\bibinfo {volume} {68}},\ \bibinfo {pages}
  {026106} (\bibinfo {year} {2003})}\BibitemShut {NoStop}%
\bibitem [{\citenamefont {Tishby}\ \emph {et~al.}(2016)\citenamefont {Tishby},
  \citenamefont {Biham},\ and\ \citenamefont {Katzav}}]{Tishby2016}%
  \BibitemOpen
  \bibfield  {author} {\bibinfo {author} {\bibfnamefont {I.}~\bibnamefont
  {Tishby}}, \bibinfo {author} {\bibfnamefont {O.}~\bibnamefont {Biham}}, \
  and\ \bibinfo {author} {\bibfnamefont {E.}~\bibnamefont {Katzav}},\
  }\href@noop {} {\bibfield  {journal} {\bibinfo  {journal} {ArXiv}\ }
  (\bibinfo {year} {2016})}\BibitemShut {NoStop}%
\bibitem [{\citenamefont {Herrero}(2005{\natexlab{a}})}]{Herrero2005-0}%
  \BibitemOpen
  \bibfield  {author} {\bibinfo {author} {\bibfnamefont {C.~P.}\ \bibnamefont
  {Herrero}},\ }\href@noop {} {\bibfield  {journal} {\bibinfo  {journal}
  {Journal of Physics A: Mathematical and General}\ }\textbf {\bibinfo {volume}
  {38}},\ \bibinfo {pages} {4349} (\bibinfo {year}
  {2005}{\natexlab{a}})}\BibitemShut {NoStop}%
\bibitem [{\citenamefont {Herrero}(2005{\natexlab{b}})}]{Herrero2005-1}%
  \BibitemOpen
  \bibfield  {author} {\bibinfo {author} {\bibfnamefont {C.}~\bibnamefont
  {Herrero}},\ }\href@noop {} {\bibfield  {journal} {\bibinfo  {journal}
  {Physical Review E}\ }\textbf {\bibinfo {volume} {71}},\ \bibinfo {pages}
  {016103} (\bibinfo {year} {2005}{\natexlab{b}})}\BibitemShut {NoStop}%
\bibitem [{\citenamefont {Fortunato}(2010)}]{Fortunato2010}%
  \BibitemOpen
  \bibfield  {author} {\bibinfo {author} {\bibfnamefont {S.}~\bibnamefont
  {Fortunato}},\ }\href@noop {} {\bibfield  {journal} {\bibinfo  {journal}
  {Physics Reports}\ }\textbf {\bibinfo {volume} {486}} (\bibinfo {year}
  {2010})}\BibitemShut {NoStop}%
\bibitem [{\citenamefont {Zhou}(2003)}]{Zhou2003}%
  \BibitemOpen
  \bibfield  {author} {\bibinfo {author} {\bibfnamefont {H.}~\bibnamefont
  {Zhou}},\ }\href@noop {} {\bibfield  {journal} {\bibinfo  {journal} {Physical
  review. E, Statistical, nonlinear, and soft matter physics}\ }\textbf
  {\bibinfo {volume} {67}},\ \bibinfo {pages} {061901} (\bibinfo {year}
  {2003})}\BibitemShut {NoStop}%
\bibitem [{\citenamefont {Rosvall}\ and\ \citenamefont
  {Bergstrom}(2008{\natexlab{a}})}]{Rosvall29012008}%
  \BibitemOpen
  \bibfield  {author} {\bibinfo {author} {\bibfnamefont {M.}~\bibnamefont
  {Rosvall}}\ and\ \bibinfo {author} {\bibfnamefont {C.~T.}\ \bibnamefont
  {Bergstrom}},\ }\href@noop {} {\bibfield  {journal} {\bibinfo  {journal}
  {Proceedings of the National Academy of Sciences}\ }\textbf {\bibinfo
  {volume} {105}},\ \bibinfo {pages} {1118} (\bibinfo {year}
  {2008}{\natexlab{a}})}\BibitemShut {NoStop}%
\bibitem [{\citenamefont {Steinhaeuser}\ and\ \citenamefont
  {Chawla}(2010)}]{Steinhaeuser2010}%
  \BibitemOpen
  \bibfield  {author} {\bibinfo {author} {\bibfnamefont {K.}~\bibnamefont
  {Steinhaeuser}}\ and\ \bibinfo {author} {\bibfnamefont {N.~V.}\ \bibnamefont
  {Chawla}},\ }\href@noop {} {\bibfield  {journal} {\bibinfo  {journal}
  {Pattern Recognition Letters}\ }\textbf {\bibinfo {volume} {31}},\ \bibinfo
  {pages} {413} (\bibinfo {year} {2010})}\BibitemShut {NoStop}%
\bibitem [{\citenamefont {Xin}\ \emph {et~al.}(2016)\citenamefont {Xin},
  \citenamefont {Xie},\ and\ \citenamefont {Yang}}]{Xin2016}%
  \BibitemOpen
  \bibfield  {author} {\bibinfo {author} {\bibfnamefont {Y.}~\bibnamefont
  {Xin}}, \bibinfo {author} {\bibfnamefont {Z.-Q.}\ \bibnamefont {Xie}}, \ and\
  \bibinfo {author} {\bibfnamefont {J.}~\bibnamefont {Yang}},\ }\href@noop {}
  {\bibfield  {journal} {\bibinfo  {journal} {Expert Systems with
  Applications}\ } (\bibinfo {year} {2016})}\BibitemShut {NoStop}%
\bibitem [{\citenamefont {Zhu}\ and\ \citenamefont {Jiang}(2016)}]{Zhu2016}%
  \BibitemOpen
  \bibfield  {author} {\bibinfo {author} {\bibfnamefont {R.}~\bibnamefont
  {Zhu}}\ and\ \bibinfo {author} {\bibfnamefont {W.}~\bibnamefont {Jiang}},\
  }\href@noop {} {\bibfield  {journal} {\bibinfo  {journal} {ArXiv}\ ,\
  \bibinfo {pages} {1}} (\bibinfo {year} {2016})}\BibitemShut {NoStop}%
\bibitem [{\citenamefont {Newman}(2006)}]{Newman06062006}%
  \BibitemOpen
  \bibfield  {author} {\bibinfo {author} {\bibfnamefont {M.~E.~J.}\
  \bibnamefont {Newman}},\ }\href@noop {} {\bibfield  {journal} {\bibinfo
  {journal} {Proceedings of the National Academy of Sciences}\ }\textbf
  {\bibinfo {volume} {103}},\ \bibinfo {pages} {8577} (\bibinfo {year}
  {2006})}\BibitemShut {NoStop}%
\bibitem [{\citenamefont {Danila}(2015)}]{Danila2015}%
  \BibitemOpen
  \bibfield  {author} {\bibinfo {author} {\bibfnamefont {B.}~\bibnamefont
  {Danila}},\ }\href@noop {} {\bibfield  {journal} {\bibinfo  {journal} {Phys.
  Rev. E}\ }\textbf {\bibinfo {volume} {022301}},\ \bibinfo {pages} {10}
  (\bibinfo {year} {2015})}\BibitemShut {NoStop}%
\bibitem [{\citenamefont {Duch}\ and\ \citenamefont
  {Arenas}(2005)}]{Arenas2005}%
  \BibitemOpen
  \bibfield  {author} {\bibinfo {author} {\bibfnamefont {J.}~\bibnamefont
  {Duch}}\ and\ \bibinfo {author} {\bibfnamefont {A.}~\bibnamefont {Arenas}},\
  }\href@noop {} {\bibfield  {journal} {\bibinfo  {journal} {Phys. Rev. E}\
  }\textbf {\bibinfo {volume} {72}},\ \bibinfo {pages} {027104} (\bibinfo
  {year} {2005})}\BibitemShut {NoStop}%
\bibitem [{\citenamefont {Newman}\ and\ \citenamefont
  {Girvan}(2004)}]{Newman2004}%
  \BibitemOpen
  \bibfield  {author} {\bibinfo {author} {\bibfnamefont {M.}~\bibnamefont
  {Newman}}\ and\ \bibinfo {author} {\bibfnamefont {M.}~\bibnamefont
  {Girvan}},\ }\href@noop {} {\bibfield  {journal} {\bibinfo  {journal}
  {Physical Review E}\ }\textbf {\bibinfo {volume} {69}},\ \bibinfo {pages}
  {026113} (\bibinfo {year} {2004})}\BibitemShut {NoStop}%
\bibitem [{\citenamefont {Newman}(2016)}]{Newman2016}%
  \BibitemOpen
  \bibfield  {author} {\bibinfo {author} {\bibfnamefont {M.~E.~J.}\
  \bibnamefont {Newman}},\ }\href@noop {} {\bibfield  {journal} {\bibinfo
  {journal} {ArXiv}\ } (\bibinfo {year} {2016})}\BibitemShut {NoStop}%
\bibitem [{\citenamefont {Defays}(1977)}]{Defays1977}%
  \BibitemOpen
  \bibfield  {author} {\bibinfo {author} {\bibfnamefont {D.}~\bibnamefont
  {Defays}},\ }\href@noop {} {\bibfield  {journal} {\bibinfo  {journal}
  {Comput. J.}\ }\textbf {\bibinfo {volume} {20}},\ \bibinfo {pages} {364}
  (\bibinfo {year} {1977})}\BibitemShut {NoStop}%
\bibitem [{\citenamefont {Day}\ and\ \citenamefont
  {Edelsbrunner}(1984)}]{Day1984}%
  \BibitemOpen
  \bibfield  {author} {\bibinfo {author} {\bibfnamefont {W.~H.}\ \bibnamefont
  {Day}}\ and\ \bibinfo {author} {\bibfnamefont {H.}~\bibnamefont
  {Edelsbrunner}},\ }\href@noop {} {\bibfield  {journal} {\bibinfo  {journal}
  {Journal of Classification}\ }\textbf {\bibinfo {volume} {1}},\ \bibinfo
  {pages} {7} (\bibinfo {year} {1984})}\BibitemShut {NoStop}%
\bibitem [{\citenamefont {Newman}\ \emph {et~al.}(2006)\citenamefont {Newman},
  \citenamefont {Barabasi},\ and\ \citenamefont {Watts}}]{Newman:2006}%
  \BibitemOpen
  \bibfield  {author} {\bibinfo {author} {\bibfnamefont {M.}~\bibnamefont
  {Newman}}, \bibinfo {author} {\bibfnamefont {A.-L.}\ \bibnamefont
  {Barabasi}}, \ and\ \bibinfo {author} {\bibfnamefont {D.~J.}\ \bibnamefont
  {Watts}},\ }\href@noop {} {\emph {\bibinfo {title} {The Structure and
  Dynamics of Networks: (Princeton Studies in Complexity)}}}\ (\bibinfo
  {publisher} {Princeton University Press},\ \bibinfo {address} {Princeton, NJ,
  USA},\ \bibinfo {year} {2006})\BibitemShut {NoStop}%
\bibitem [{\citenamefont {W}\ \emph {et~al.}(2016)\citenamefont {W},
  \citenamefont {X}, \citenamefont {M},\ and\ \citenamefont {X}}]{W2016}%
  \BibitemOpen
  \bibfield  {author} {\bibinfo {author} {\bibfnamefont {L.}~\bibnamefont {W}},
  \bibinfo {author} {\bibfnamefont {J.}~\bibnamefont {X}}, \bibinfo {author}
  {\bibfnamefont {P.}~\bibnamefont {M}}, \ and\ \bibinfo {author}
  {\bibfnamefont {W.}~\bibnamefont {X}},\ }\href@noop {} {\bibfield  {journal}
  {\bibinfo  {journal} {Nature Publishing Group}\ ,\ \bibinfo {pages} {22470}}
  (\bibinfo {year} {2016})}\BibitemShut {NoStop}%
\bibitem [{\citenamefont {Abdi}\ and\ \citenamefont
  {Williams}(2010)}]{Abdi2010}%
  \BibitemOpen
  \bibfield  {author} {\bibinfo {author} {\bibfnamefont {H.}~\bibnamefont
  {Abdi}}\ and\ \bibinfo {author} {\bibfnamefont {L.~J.}\ \bibnamefont
  {Williams}},\ }\href@noop {} {\bibfield  {journal} {\bibinfo  {journal}
  {Wiley Interdisciplinary Reviews: Computational Statistics}\ }\textbf
  {\bibinfo {volume} {2}},\ \bibinfo {pages} {433} (\bibinfo {year}
  {2010})}\BibitemShut {NoStop}%
\bibitem [{\citenamefont {Bellman}(1957)}]{Bellman1957}%
  \BibitemOpen
  \bibfield  {author} {\bibinfo {author} {\bibfnamefont {R.}~\bibnamefont
  {Bellman}},\ }\href@noop {} {\emph {\bibinfo {title} {Dynamic Programming}}}\
  (\bibinfo  {publisher} {Princeton University Press},\ \bibinfo {year}
  {1957})\BibitemShut {NoStop}%
\bibitem [{\citenamefont {Kindt}\ and\ \citenamefont {Coe}(2005)}]{Kindt2005}%
  \BibitemOpen
  \bibfield  {author} {\bibinfo {author} {\bibfnamefont {R.}~\bibnamefont
  {Kindt}}\ and\ \bibinfo {author} {\bibfnamefont {R.}~\bibnamefont {Coe}},\
  }\href@noop {} {\bibfield  {journal} {\bibinfo  {journal} {Tree diversity
  analysis}\ ,\ \bibinfo {pages} {123}} (\bibinfo {year} {2005})}\BibitemShut
  {NoStop}%
\bibitem [{\citenamefont {Lancichinetti}\ and\ \citenamefont
  {Fortunato}(2011)}]{Fortunato2011}%
  \BibitemOpen
  \bibfield  {author} {\bibinfo {author} {\bibfnamefont {A.}~\bibnamefont
  {Lancichinetti}}\ and\ \bibinfo {author} {\bibfnamefont {S.}~\bibnamefont
  {Fortunato}},\ }\href@noop {} {\bibfield  {journal} {\bibinfo  {journal}
  {Phys. Rev. E}\ }\textbf {\bibinfo {volume} {84}},\ \bibinfo {pages} {066122}
  (\bibinfo {year} {2011})}\BibitemShut {NoStop}%
\bibitem [{\citenamefont {Fortunato}\ and\ \citenamefont
  {Barthélemy}(2007)}]{Fortunato2012}%
  \BibitemOpen
  \bibfield  {author} {\bibinfo {author} {\bibfnamefont {S.}~\bibnamefont
  {Fortunato}}\ and\ \bibinfo {author} {\bibfnamefont {M.}~\bibnamefont
  {Barthélemy}},\ }\href@noop {} {\bibfield  {journal} {\bibinfo  {journal}
  {Proceedings of the National Academy of Sciences}\ }\textbf {\bibinfo
  {volume} {104}},\ \bibinfo {pages} {36} (\bibinfo {year} {2007})}\BibitemShut
  {NoStop}%
\bibitem [{\citenamefont {Fortunato}\ and\ \citenamefont
  {Hric}(2016)}]{Fortunato2016}%
  \BibitemOpen
  \bibfield  {author} {\bibinfo {author} {\bibfnamefont {S.}~\bibnamefont
  {Fortunato}}\ and\ \bibinfo {author} {\bibfnamefont {D.}~\bibnamefont
  {Hric}},\ }\href@noop {} {\bibfield  {journal} {\bibinfo  {journal} {Physics
  Reports}\ }\textbf {\bibinfo {volume} {659}},\ \bibinfo {pages} {1} (\bibinfo
  {year} {2016})}\BibitemShut {NoStop}%
\bibitem [{\citenamefont {Newman}\ and\ \citenamefont
  {Girvan}(2002)}]{Girvan2002}%
  \BibitemOpen
  \bibfield  {author} {\bibinfo {author} {\bibfnamefont {M.}~\bibnamefont
  {Newman}}\ and\ \bibinfo {author} {\bibfnamefont {M.}~\bibnamefont
  {Girvan}},\ }\href@noop {} {\bibfield  {journal} {\bibinfo  {journal}
  {Proceedings of the National Academy of Sciences of the United States of
  America}\ }\textbf {\bibinfo {volume} {99}},\ \bibinfo {pages} {7821}
  (\bibinfo {year} {2002})}\BibitemShut {NoStop}%
\bibitem [{\citenamefont {Lancichinetti}\ \emph {et~al.}(2008)\citenamefont
  {Lancichinetti}, \citenamefont {Fortunato},\ and\ \citenamefont
  {Radicchi}}]{Lancichinetti2008}%
  \BibitemOpen
  \bibfield  {author} {\bibinfo {author} {\bibfnamefont {A.}~\bibnamefont
  {Lancichinetti}}, \bibinfo {author} {\bibfnamefont {S.}~\bibnamefont
  {Fortunato}}, \ and\ \bibinfo {author} {\bibfnamefont {F.}~\bibnamefont
  {Radicchi}},\ }\href@noop {} {\bibfield  {journal} {\bibinfo  {journal}
  {Physical Review E}\ }\textbf {\bibinfo {volume} {78}},\ \bibinfo {pages} {1}
  (\bibinfo {year} {2008})}\BibitemShut {NoStop}%
\bibitem [{\citenamefont {Zachary}(1977)}]{Zachary1977}%
  \BibitemOpen
  \bibfield  {author} {\bibinfo {author} {\bibfnamefont {W.}~\bibnamefont
  {Zachary}},\ }\href@noop {} {\bibfield  {journal} {\bibinfo  {journal}
  {Journal of anthropological research}\ }\textbf {\bibinfo {volume} {33}},\
  \bibinfo {pages} {452} (\bibinfo {year} {1977})}\BibitemShut {NoStop}%
\bibitem [{\citenamefont {Lusseau}(2003)}]{Lusseau2003}%
  \BibitemOpen
  \bibfield  {author} {\bibinfo {author} {\bibfnamefont {D.}~\bibnamefont
  {Lusseau}},\ }\href@noop {} {\bibfield  {journal} {\bibinfo  {journal}
  {Proceedings. Biological sciences / The Royal Society}\ }\textbf {\bibinfo
  {volume} {270 Suppl}},\ \bibinfo {pages} {S186} (\bibinfo {year}
  {2003})}\BibitemShut {NoStop}%
\bibitem [{\citenamefont {Knuth}(1993)}]{knuth1993}%
  \BibitemOpen
  \bibfield  {author} {\bibinfo {author} {\bibfnamefont {D.~E.}\ \bibnamefont
  {Knuth}},\ }\href@noop {} {\emph {\bibinfo {title} {The {Stanford}
  {GraphBase}: A Platform for Combinatorial Computing}}},\ Vol.~\bibinfo
  {volume} {37}\ (\bibinfo  {publisher} {Addison-Wesley Reading},\ \bibinfo
  {year} {1993})\BibitemShut {NoStop}%
\bibitem [{\citenamefont {Gleiser}\ and\ \citenamefont
  {Danon}(2003)}]{Gleiser2003}%
  \BibitemOpen
  \bibfield  {author} {\bibinfo {author} {\bibfnamefont {P.}~\bibnamefont
  {Gleiser}}\ and\ \bibinfo {author} {\bibfnamefont {L.}~\bibnamefont
  {Danon}},\ }\href@noop {} {\bibfield  {journal} {\bibinfo  {journal}
  {Advances in complex systems}\ }\textbf {\bibinfo {volume} {6}},\ \bibinfo
  {pages} {12} (\bibinfo {year} {2003})}\BibitemShut {NoStop}%
\bibitem [{\citenamefont {White}\ \emph {et~al.}(1986)\citenamefont {White},
  \citenamefont {Southgate}, \citenamefont {Thomson},\ and\ \citenamefont
  {Brenner}}]{White1986}%
  \BibitemOpen
  \bibfield  {author} {\bibinfo {author} {\bibfnamefont {J.~G.}\ \bibnamefont
  {White}}, \bibinfo {author} {\bibfnamefont {E.}~\bibnamefont {Southgate}},
  \bibinfo {author} {\bibfnamefont {J.~N.}\ \bibnamefont {Thomson}}, \ and\
  \bibinfo {author} {\bibfnamefont {S.}~\bibnamefont {Brenner}},\ }\href@noop
  {} {\bibfield  {journal} {\bibinfo  {journal} {Philosophical Transactions of
  the Royal Society of London B: Biological Sciences}\ }\textbf {\bibinfo
  {volume} {314}},\ \bibinfo {pages} {1} (\bibinfo {year} {1986})}\BibitemShut
  {NoStop}%
\bibitem [{\citenamefont {Watts}\ and\ \citenamefont
  {Strogatz}(1998)}]{WattsDJ1998}%
  \BibitemOpen
  \bibfield  {author} {\bibinfo {author} {\bibfnamefont {D.~J.}\ \bibnamefont
  {Watts}}\ and\ \bibinfo {author} {\bibfnamefont {S.~H.}\ \bibnamefont
  {Strogatz}},\ }\href@noop {} {\bibfield  {journal} {\bibinfo  {journal}
  {Nature}\ ,\ \bibinfo {pages} {440}} (\bibinfo {year} {1998})}\BibitemShut
  {NoStop}%
\bibitem [{\citenamefont {Clauset}\ \emph {et~al.}(2004)\citenamefont
  {Clauset}, \citenamefont {Newman},\ and\ \citenamefont
  {Moore}}]{Clauset2004}%
  \BibitemOpen
  \bibfield  {author} {\bibinfo {author} {\bibfnamefont {A.}~\bibnamefont
  {Clauset}}, \bibinfo {author} {\bibfnamefont {M.~E.~J.}\ \bibnamefont
  {Newman}}, \ and\ \bibinfo {author} {\bibfnamefont {C.}~\bibnamefont
  {Moore}},\ }\href@noop {} {\bibfield  {journal} {\bibinfo  {journal}
  {Physical Review E}\ }\textbf {\bibinfo {volume} {70}},\ \bibinfo {pages} {1}
  (\bibinfo {year} {2004})}\BibitemShut {NoStop}%
\bibitem [{\citenamefont {Pons}\ and\ \citenamefont {Latapy}(2005)}]{Pons2005}%
  \BibitemOpen
  \bibfield  {author} {\bibinfo {author} {\bibfnamefont {P.}~\bibnamefont
  {Pons}}\ and\ \bibinfo {author} {\bibfnamefont {M.}~\bibnamefont {Latapy}},\
  }\href@noop {} {\bibfield  {journal} {\bibinfo  {journal} {Journal of Graph
  Algorithms and Applications}\ }\textbf {\bibinfo {volume} {10}},\ \bibinfo
  {pages} {191} (\bibinfo {year} {2005})}\BibitemShut {NoStop}%
\bibitem [{\citenamefont {Rosvall}\ and\ \citenamefont
  {Bergstrom}(2008{\natexlab{b}})}]{Rosvall2008}%
  \BibitemOpen
  \bibfield  {author} {\bibinfo {author} {\bibfnamefont {M.}~\bibnamefont
  {Rosvall}}\ and\ \bibinfo {author} {\bibfnamefont {C.~T.}\ \bibnamefont
  {Bergstrom}},\ }\href {\doibase 10.1073/pnas.0706851105} {\bibfield
  {journal} {\bibinfo  {journal} {Proceedings of the National Academy of
  Sciences}\ }\textbf {\bibinfo {volume} {105}},\ \bibinfo {pages} {1118}
  (\bibinfo {year} {2008}{\natexlab{b}})}\BibitemShut {NoStop}%
\bibitem [{\citenamefont {Lancichinetti}\ \emph {et~al.}(2009)\citenamefont
  {Lancichinetti}, \citenamefont {Fortunato},\ and\ \citenamefont
  {Kert{\'{e}}sz}}]{Lancichinetti2009}%
  \BibitemOpen
  \bibfield  {author} {\bibinfo {author} {\bibfnamefont {A.}~\bibnamefont
  {Lancichinetti}}, \bibinfo {author} {\bibfnamefont {S.}~\bibnamefont
  {Fortunato}}, \ and\ \bibinfo {author} {\bibfnamefont {J.}~\bibnamefont
  {Kert{\'{e}}sz}},\ }\href@noop {} {\bibfield  {journal} {\bibinfo  {journal}
  {New Journal of Physics}\ }\textbf {\bibinfo {volume} {11}},\ \bibinfo
  {pages} {1} (\bibinfo {year} {2009})}\BibitemShut {NoStop}%
\bibitem [{\citenamefont {Donetti}\ and\ \citenamefont
  {Munoz}(2004)}]{Donetti2004}%
  \BibitemOpen
  \bibfield  {author} {\bibinfo {author} {\bibfnamefont {L.}~\bibnamefont
  {Donetti}}\ and\ \bibinfo {author} {\bibfnamefont {M.~A.}\ \bibnamefont
  {Munoz}},\ }\href@noop {} {\bibfield  {journal} {\bibinfo  {journal} {J.
  Stat. Mech.}\ }\textbf {\bibinfo {volume} {P10012}} (\bibinfo {year}
  {2004})}\BibitemShut {NoStop}%
\bibitem [{\citenamefont {Shen}\ \emph {et~al.}(2009)\citenamefont {Shen},
  \citenamefont {Cheng},\ and\ \citenamefont {Guo}}]{Hua-Wei2009}%
  \BibitemOpen
  \bibfield  {author} {\bibinfo {author} {\bibfnamefont {H.-W.}\ \bibnamefont
  {Shen}}, \bibinfo {author} {\bibfnamefont {X.-Q.}\ \bibnamefont {Cheng}}, \
  and\ \bibinfo {author} {\bibfnamefont {J.-F.}\ \bibnamefont {Guo}},\
  }\href@noop {} {\bibfield  {journal} {\bibinfo  {journal} {Journal of
  Statistical Mechanics: Theory and Experiment}\ }\textbf {\bibinfo {volume}
  {2009}},\ \bibinfo {pages} {P07042} (\bibinfo {year} {2009})}\BibitemShut
  {NoStop}%
\bibitem [{\citenamefont {Zhang}(2012)}]{Zhang2012}%
  \BibitemOpen
  \bibfield  {author} {\bibinfo {author} {\bibfnamefont {S.}~\bibnamefont
  {Zhang}},\ }\href@noop {} {\bibfield  {journal} {\bibinfo  {journal}
  {TheScientificWorldJournal}\ }\textbf {\bibinfo {volume} {2012}},\ \bibinfo
  {pages} {523706} (\bibinfo {year} {2012})}\BibitemShut {NoStop}%
\end{thebibliography}%

\end{document}